\documentstyle[epsf,psfig,prl,multicol,aps]{revtex}
\begin{document}
\title{Vibrational Modes in LiBC: Theory Compared with Experiment}

\author{J. M. An,$^{a,b}$, H. Rosner,$^b$,
        S. Y. Savrasov$^c$, and W. E. Pickett,$^b$}
\address{$^a$Lawrence Berkeley National Laboratory, Berkeley CA 94720\\
     $^b$Physics Department, University of California, Davis CA 95616\\
     $^c$Department of Physics, New Jersey Institute of Technology, 
               Newark NJ 07102}

\date{\today}
\maketitle
\begin{abstract}
The search for other superconductors in the MgB$_2$ class currently is
focussed on Li$_{1-x}$BC, which when hole-doped (concentration $x$)
should be a metal with the potential to be a better superconductor than
MgB$_2$.  Here we present the calculated phonon spectrum of the parent
semiconductor LiBC.  The calculated Raman-active modes are in excellent
agreement with a recent observation, and comparison of calculated IR-active
modes with a recent report provides a prediction of the LO--TO splitting
for these four modes, which is small for the B-C bond stretching mode
at $\sim$1200 cm$^{-1}$, but large for clearly resolved modes at
540 cm$^{-1}$ and 620 cm$^{-1}$.
\end{abstract}


\begin{multicols}{2}
\section{Introduction}
The discovery of MgB$_2$ with T$_c \sim$ 40 
K \cite{akimitsu} has led to the search of related materials that may
show superconductivity.  An
important feature required for strong electron-phonon coupling 
in this class of materials is
that the B-C bonding $\sigma$ bands need to
be partially unfilled as in MgB$_2$, which is not the case in 
AlB$_2$.\cite{bohnen}
Several isostructural transition metal diborides exist,\cite{tab2} but their
electronic structures are entirely different and not conducive to
high temperature superconductivity.

\vskip 4mm
\begin{figure}[bt]
\begin{center}
\psfig{figure=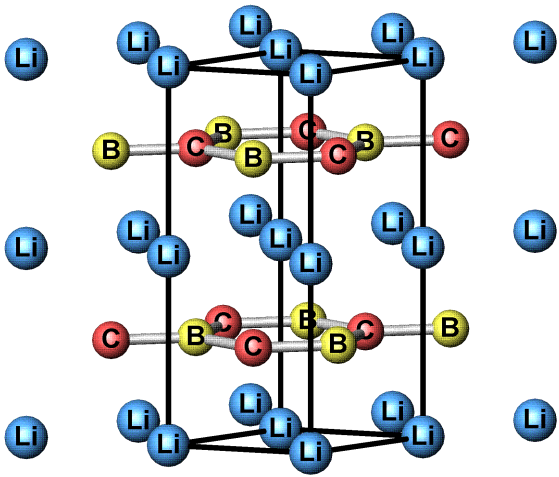,width=7.5cm,angle=-0}
\end{center}
\caption{
Crystal structure of LiBC.  The graphene structure of the B-C layers is 
a direct generalization of the B layers of MgB$_2$, and the Li atoms
reside on the Mg site of MgB$_2$.  The unit cell contains two 
formula units because B and C alternate in position along the $\hat c$
axis ({\i.e.} the two B-C layers are connected by a 120$^{\circ}$ 
screw axis).
}
\label{fig1}
\end{figure}

Some of the present authors have proposed\cite{libcucd}
that hole-doped LiBC, which was reported by W\"orle
{\it et al.},\cite{worle} should be a member of the MgB$_2$ class of
superconductors and would have an even higher T$_c$ if it is doped
to $x > 0.3$.\cite{libcucd,predict}  Several other groups have now reported
the undoped, semiconducting LiBC compound\cite{india,dresden,IR,raman}
and have reported characterization by x-ray diffraction, infrared (IR)
absorption, and Raman spectroscopy.

The theoretical work since January 
2001\cite{bohnen,jan,kortus,kong,yildirim,liu} has clarified much about
the cause of the high T$_c$ in MgB$_2$: 
strongly covalent-bonding states, normally
fully occupied, are driven to the Fermi level (E$_F$) by the chemistry of
MgB$_2$, and the resulting hole carriers are exceedingly strongly coupled
to the bond stretching modes.  This ``covalent'' coupling is what drives
the critical temperature from zero (or near) to 40 K.

\section{Method of Calculation}
The phonon energies and
EP matrix elements have been obtained from linear
response theory,\cite{linear} as implemented in Savrasov's
full-potential linear
muffin-tin orbital code.\cite{Sav,LMTO}  A double-$\kappa$ basis was
sufficient, since the atoms only have open $s$ and $p$ shells.
A dense grid of Q points was
chosen (a 16,16,4 grid giving 90 Q points in the irreducible wedge).
For k-space integration a finer 32,32,8 grid was used, together
with an adaptive tetrahedron integration scheme.  The code was used
previously for MgB$_2$ by Andersen's group.\cite{kong} 
These Q meshes and k-point meshes were chosen to deal with Fermi
surface effects in the hole-doped (hence metallic) material, and should
be more than sufficient for the semiconducting phase that 
we discuss here.  The 
effects of doping on the phonon spectrum are reported 
elsewhere.\cite{predict}

The frozen phonons were calculated using one of two methods.
The linearized augmented
plane method as implemented in the WIEN97 package\cite{wien} and
the full potential local orbital (FPLO) code\cite{eschrig89,koepernik99}
were used.  Calculational
details were described previously.\cite{libcucd,predict}
The masses used were, in a.m.u.: Li, 6.94; B, 10.81; C, 12.01, and
the experimental lattice constants $a$=2.75\AA, $c$=7.058\AA~were used. 

\section{Calculated Phonon Frequencies at Q=0}
\begin{figure}[bt]
\begin{center}
\psfig{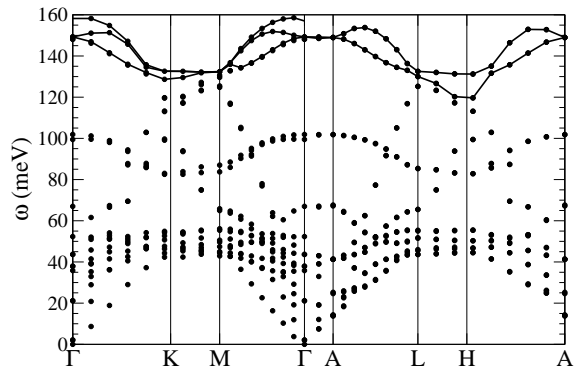}
\end{center}
\caption{
Phonon spectrum of semiconducting LiBC, calculated using linear response
methods as described in the text.  The bond-stretching modes are
connected by solid lines.
}
\label{fig2}
\end{figure}

The results for all eighteen branches along high symmetry directions in
the hexagonal Brillouin zone are shown in Fig.~\ref{fig2}.  We 
concentrate here on the Q=0 modes that are most accessible to
experimental probes.
According to the code\cite{stokes} SMODES.1.2.4 that calculates the 
eigenmodes for an arbitrary crystal structure and Q point, the $\Gamma$
point modes in LiBC (P6$_3$/$mmc$, \#194 in the International Tables)
decompose as 3A$_{2u}$+2B$_{2g}$+1B$_{1u}$ (6 modes)
polarized along the $\hat c$
axis, and 3E$_{1u}$+2E$_{2g}$+E$_{2u}$ (12 modes)
polarized in the basal plane.
The 2E$_{2g}$ modes are Raman active, while IR activity involves the
2A$_{2u}$ and 2E$_{1u}$ modes.\cite{IR}
(A different convention for designating the B$_{1g}$ and B$_{2g}$ symmetry
labels was used in Ref.~\cite{raman}.  We use the convention used by the 
SMODES code.)
We go through the modes calculated using linear response theory
individually in increasing frequency (given in cm$^{-1}$).

\begin{itemize}
\item $\omega$=0:  A$_{2u}$ and E$_{1u}$, acoustic modes.
\item $\omega$=171: E$_{2g}$, B-C layers sliding against each
  other 
\item $\omega$=289: B$_{2g}$, B-C layers beating against each other,
  motion along the $\hat c$ axis 
\item $\omega$=306: E$_{2u}$, Li layers sliding against each other
\item $\omega$=352: E$_{1u}$, Li layers sliding against the B-C 
   layers 
\item $\omega$=422: A$_{2u}$, Li layers beating against the B-C layers,
  motion along the $\hat c$ axis
\item $\omega$=540: B$_{1u}$, Li layers beating against each other,
  motion along the $\hat c$ axis
\item $\omega$=802: A$_{2u}$, B-C puckering mode, all B atoms 
  move oppositely
  to all C atoms, Li sites become inequivalent
\item $\omega$=821: B$_{2g}$, B-C puckering mode, B moves with C atoms
  above/below it, Li sites remain equivalent
\item $\omega$=1194: E$_{1u}$, B-C bond stretching mode, the two
  layers are out-of-phase
\item $\omega$=1204: E$_{2g}$, B-C bond stretching mode, layers in phase
\end{itemize}
As expected, the lower frequency modes $\omega <$ 600 cm$^{-1}$ involve
``rigid'' layer displacements (B and C displacements differ somewhat,
but are parallel for these modes).  The puckering (B-C bond bending) 
modes lie at 811$\pm$10 cm$^{-1}$ = 100 meV, and both in-phase and 
out-of-phase bond-stretching modes are at very high frequency
(1200 cm$^{-1}$ = 149 meV).

\section{Checks Using the Frozen Phonon Method}
During the progress of our studies we have made various checks on the
results of the linear response calculations by using the (reliable but
tedious) frozen phonon method for selected modes.  Here we briefly 
describe these results, which serve to verify the linear response results
and provide some indication of the precision of our methods. 
Each of the methods is subject to its own numerical approximations,
and hence uncertainties.  For example, the frozen phonon method requires
the fitting of a discrete set of E(u) (energy versus displacement) values
to a functional form, followed by use of the quadratic term, whereas
the linear response method does not, but is more sensitive to choice of
basis set, for example.

Symmetry lowering from the ideal P$6_3/mmc$ space group observed in
Raman scattering spectra\cite{raman} suggested a puckering of the B-C
rings and change of crystal symmetry to the P$\bar 3m1$ space group, 
which would imply an instability of the flat B-C layer to such a
puckering displacement.  Calculation of the energy change versus displacement 
for this buckling mode indicated not only no instability, but a 
rather hard phonon with frequency of 788 cm$^{-1}$, within 2.5\% of
the linear response result of 802 cm$^{-1}$.  
This result is consistent with the tight-binding model of Ramirez
{\it et al.}\cite{ramirez} for a single B-C layer coupled to ionic
charges representing the rest of the crystal, which gave a stable 
flat (unpuckered) layer.
Another 
possibility that may account for this observed symmetry lowering would be
layer stacking faults (B above B and C above C).  Such faults would provide
another distinct type of Li site, in much the same way that occurs for the
(proposed) frozen puckering mode.  Further experimental work will be 
necessary to clarify this question.

Frozen phonon evaluations of the E$_{2g}$ bond-stretching mode were
carried out using both the FPLO and LAPW codes.  The results (1167 cm$^{-1}$
and 1145 cm$^{-1}$, respectively) are 4\% lower than the linear response
result.  However, some of this discrepancy is due to neglect of 
coupling of this mode with the lower frequency E$_{2g}$ mode.  (The
``frozen phonon'' we chose assumed the B and C displacement amplitudes
 were inversely proportional to their masses.)  This neglected coupling 
would increase the frequency, improving agreement with the linear
response value.

\section{Comparison with Experiment}

Raman active E$_{2g}$ modes are observed at 170 and 
1176 cm$^{-1}$.\cite{raman}
The harder mode is surely the bond-stretching mode, calculated to lie
at 1204 (linear response) and 1145, 1167 cm$^{-1}$ by the two independent
FP calculations discussed above (which, as noted there, are known to be
slight underestimates).  Thus the linear response value is too large
by 2\%.
The 170 mode arises from
the B-C ``shear'' mode (Li layers are fixed)\cite{IR} and is
indistinguishable from the calculated 171 cm$^{-1}$.

The excellent agreement for the bond-stretching mode implies that
anharmonicity must be negligible for LiBC.  Several groups have 
calculated the harmonic frequency of the Raman-active mode in MgB$_2$,
finding results in the range 490 -- 550 cm$^{-1}$.  The peak in the
observed Raman spectrum occurs at 600 cm$^{-1}$, which has been 
interpreted to mean that
anharmonicity of this mode in MgB$_2$ is 
considerable.\cite{yildirim,liu}  
Boeri {\it et al.} have
demonstrated that this anharmonicity is due to the metallic nature,
and to the proximity of the Fermi level to the $\sigma$ band 
edge.\cite{boeri}
Hence, appreciable anharmonicity is not expected to be present in the
undoped LiBC material.  We have found, in hole-doped Li$_{1-x}$BC,
that this anharmonicity again arises, and this work will be reported
elsewhere.

IR absorption measurements, performed with samples and geometry such that 
all modes are likely to be seen, identify broad but reasonably well 
defined peaks centered at 540, 620, and 1180 cm$^{-1}$, and a shoulder
in the 680-800 cm$^{-1}$ region that could be fit to a IR mode around
700 cm$^{-1}$.\cite{IR}  Comparison of calculations with IR measurements 
is problematic unless the dynamical effective charges Z$^*$ are calculated,
which we have not yet done.  All of our calculations neglect the 
accompanying macroscopic electric field which gives the LO-TO
splitting.  Since the calculations agree so well with data for the
Raman modes,
we will interpret the difference $\omega_{exp} - \omega_{calc}$ as the
LO-TO splitting for the mode, and we will only know of difficulties if
this difference is negative.

The observed 1180 cm$^{-1}$ mode is surely the E$_{1u}$ B-C bond stretching 
mode, calculated to lie at 1194 cm$^{-1}$.  
The agreement implies that the LO-TO splitting for this mode
is very small (it cannot be negative, of course).  
The best identification of the experimental ``shoulder''
mode at $\sim 700$ cm$^{-1}$ is with the calculated A$_{2u}$ B-C
puckering mode at 802 cm$^{-1}$, which indicates a problem either
with the identification of this shoulder with an IR-active mode,
or a problem with the calculation; further experimental study will
be required to resolve this discrepancy.
The next lower calculated mode is the E$_{1u}$ Li sliding mode,
$\omega_{calc}=306, \omega_{exp}=620$ (implying an unexpectedly large
LO-TO splitting of 315 cm$^{-1}$), since it would imply a large Z$^*$
for the Li ion.  The last, and lowest, mode is
the B-C layer beating mode, $\omega_{calc}=289, \omega_{exp}=540$.
This large LO-TO splitting of 250 cm$^{-1}$ implies large 
Z$^*$ for the B and/or C atoms for $\hat c$ axis displacement.   

\section{Summary}
The linear response phonon calculations are in quite close agreement with
the Raman scattering identifications of Hlinka {\it et al.}\cite{raman}
seen in $xy$ polarization.  The symmetry lowering inferred from 
scattering in $xx$ polarization remains unclear.  Comparison of the
calculations with the IR microscope results of 
Pronin {\it et al.}\cite{IR} for the lower frequency
IR modes suggest large Z$^*$ for the Li and B (and perhaps C)
ions for $\hat c$ axis displacements, while the lack of any significant
LO-TO splitting for the bond-stretching mode suggests negligible Z$^*$
for $a-b$ plane displacement of B and C.

We acknowledge helpful communications with A. Loidl and J. Hlinka.
This work was supported by National Science Foundation Grant
DMR-0114818, and by the Deutscher Akademischer Austauschdienst.



\end{multicols}
\end{document}